\documentclass{sig-alternate-2013}
\newfont{\mycrnotice}{ptmr8t at 7pt}
\newfont{\myconfname}{ptmri8t at 7pt}
\let\crnotice\mycrnotice%
\let\confname\myconfname%

\permission{}
\conferenceinfo{CIKM'13,}{Oct. 27--Nov. 1, 2013, San Francisco, CA, USA.\\http://dx.doi.org/10.1145/2505515.2505541}
\copyrightetc{}
\crdata{http://dx.doi.org/10.1145/2505515.2505541}

\usepackage{booktabs}
\usepackage{graphicx,subfigure}
\usepackage{algorithm} 
\usepackage{algorithmic} 
\usepackage{multirow} 
\usepackage{amsmath}
\usepackage{xcolor}
\usepackage{tikz}
\usepackage{graphicx}
\usepackage{subfigure}
\usetikzlibrary{arrows}
\tikzstyle{block}=[draw opacity=0.7,line width=1.4cm]
\begin{document}
%

\title{StaticGreedy: Solving the Scalability-Accuracy Dilemma in Influence Maximization}
%
%
%
%
%

\numberofauthors{1} 
%
\author{
%
%
\alignauthor
Suqi Cheng, Huawei Shen, Junming Huang, Guoqing Zhang, Xueqi Cheng\\
       \affaddr{Institute of Computing Technology, Chinese Academy of Sciences, Beijing, China}\\
       \email{\{chengsuqi, shenhuawei, huangjunming, gqzhang, cxq\}@ict.ac.cn}
}

\maketitle
\begin{abstract}
  Influence maximization, defined as a problem of finding a set of seed nodes to trigger a maximized spread of influence, is crucial to viral marketing on social networks. For practical viral marketing on large scale social networks, it is required that influence maximization algorithms should have both guaranteed accuracy and high scalability. However, existing algorithms suffer a scalability-accuracy dilemma: conventional greedy algorithms guarantee the accuracy with expensive computation, while the scalable heuristic algorithms suffer from unstable accuracy.

  In this paper, we focus on solving this scalability-accuracy dilemma. We point out that the essential reason of the dilemma is the surprising fact that the submodularity, a key requirement of the objective function for a greedy algorithm to approximate the optimum, is not guaranteed in all conventional greedy algorithms in the literature of influence maximization. Therefore a greedy algorithm has to afford a huge number of Monte Carlo simulations to reduce the pain caused by unguaranteed submodularity. Motivated by this critical finding, we propose a static greedy algorithm, named \textbf{StaticGreedy}, to strictly guarantee the submodularity of influence spread function during the seed selection process. The proposed algorithm makes the computational expense dramatically reduced by two orders of magnitude without loss of accuracy. Moreover, we propose a dynamical update strategy which can speed up the StaticGreedy algorithm by 2-7 times on large scale social networks.


\end{abstract}



\section{Introduction}

We are witnessing the increasing prosperity of online social network sites and social media sites, where people are connected by heterogeneous social relationships. These online social networks provide convenient platforms for information dissemination and marketing campaign, allowing ideas and behaviors to flow along the social relationships in the effective word-of-mouth manner. Many companies have made efforts to popularize or promote their brands or products on online social networks by launching campaigns akin to viral marketing. The success of viral marketing is rooted in the interpersonal influence, which has been empirically studied in various contexts~\cite{Domingos2001,Richardson2002,Kimura2006,Huang2010,Huang2012,Li2012,Ye2012,Bao2013}.

Influence maximization, formulated as a discrete optimization problem by Kempe et al.~\cite{Kempe2003}, is a fundamental problem for viral marketing. It aims to find a fixed-size set of seed nodes, which can influence the maximum number of nodes, generally referred to as \emph{influence spread}. The solution of the influence maximization problem is closely related to information spread models, which are used to model the process of influence spread. Two commonly-used models are the independent cascade model and the linear threshold model. Kempe et al.~\cite{Kempe2003} proved the influence maximization problem is NP-hard with either model, and proposed a greedy algorithm to approximate the optimal solution within a factor of $(1-1/\mathrm{e}-\epsilon)$, where $\epsilon$ depends on the accuracy of influence spread estimation. Since no algorithm can efficiently estimate the exact influence spread of a given seed set on typically sized networks~\cite{Chen2010,Chen2010b}, Monte Carlo approach is usually used to provide an approximation, resulting a small positive error $\epsilon$. 
%

Unfortunately, the greedy algorithm proposed by Kempe et al. (referred to as \textbf{GeneralGreedy} in this paper) suffers severe scalability problem, i.e., it relies on a huge number of Monte Carlo simulations to achieve a fair solution, which results in an unaffordable computation on large-scale social networks. To overcome this problem, many efforts have been made to explore a more scalable greedy algorithm along two directions~\cite{Leskovec2007,Chen2009,Kimura2010,Wang2010,Narayanam2011,Jiang2011,Goyal2011}. On one direction, researchers insisted on Monte Carlo simulations and reduced the number of trials that need Monte Carlo simulations to estimate the influence spreads of node sets. For example, a ``lazy-forward'' strategy was proposed to effectively reduce the number of candidate nodes~\cite{Leskovec2007}. However, the reduction in computational expense was limited since a large number of Monte Carlo simulations were still needed in every single estimation to guarantee the final accuracy. On the other direction, various heuristics were proposed to efficiently estimate influence spreads instead of Monte Carlo simulations. In a representative work,
the maximum influence paths between every pair of nodes are used to approximately compute the influence propagation~\cite{Chen2010}. However, the gain in scalability is obtained with the pain of unguaranteed accuracy. In a word, existing influence maximization algorithms suffer the scalability-accuracy dilemma.

This paper focuses on resolving the scalability-accuracy dilemma of influence maximization with respect to the independent cascade model. We analyze the essential cause of the scalability-accuracy dilemma, and then propose a static greedy algorithm to combat it. Moreover, we further improve the scalability of the proposed algorithm by a dynamic update strategy. The contributions of this paper are summarized as follows:

\begin{itemize}
\item
  We point out the cause of the expensive computation is that the submodularity is not strictly guaranteed in existing greedy algorithms. 
  Failing to strictly guarantee the submodularity, one needs to run a large number of Monte Carlo simulations to approximately guarantee the submodularity, which results in an unaffordable computational expense. This critical finding renews our knowledge about greedy algorithms for influence maximization and opens a door to resolve the scalability-accuracy dilemma.

\item
  We propose a static greedy algorithm to strictly guarantee the submodularity property of influence spread by reusing the results of Monte Carlo simulation during the whole process of greedy selection. The algorithm dramatically reduces its computational expense by two orders of magnitude without loss of accuracy.

\item

  We further speed up our algorithm by dynamically updating the marginal gain of the candidate nodes. This updating strategy, taking the advantage of static results of Monte Carlo simulations, makes our algorithm $2-7$ times faster than the StaticGreedy algorithm optimized by CELF. The improved algorithm has a speed comparable with the most scalable heuristic algorithm.



\end{itemize}

\section{Related Work}

Influence maximization was first studied by Domingos and Richardson from the algorithmic perspective~\cite{Domingos2001,Richardson2002}, and was then formulated as a discrete optimization problem by Kempe et al.~\cite{Kempe2003}. They also proposed a greedy algorithm, with the accuracy guaranteed by the monotonicity and submodularity properties of the objective function of influence maximization problem. However, this greedy algorithm is inefficient and not scalable to large scale social networks.

Thus, several studies were devoted to optimize Kepme's greedy algorithm without affecting its guaranteed accuracy. Leskovec et al.~\cite{Leskovec2007} proposed the ``cost-effective lazy forward'' strategy, namely CELF, for selecting new seed nodes by further exploiting the submodularity property of influence maximization. The CELF strategy can greatly reduce the number of evaluations on the influence spread of nodes. This strategy was further improved to a CELF++ strategy~\cite{Goyal2011}, which simultaneously calculates the influence spread for two successive iterations of greedy algorithm. NewGreedy algorithm~\cite{Chen2009} reuses the results of Monte Carlo simulations to estimate the influence spread for all candidate nodes in the same iteration. It has been further developed into MixedGreedy algorithm to integrate the advantages of both the CELF strategy and the NewGreedy algorithm.

Unfortunately, those improved greedy algorithms are still inefficient for involving too many Monte Carlo simulations for influence spread estimation. Hence, several heuristics for the independent cascade model were proposed to improve the scalability of greedy algorithm by simplifying influence spread estimation. Chen et al.~\cite{Chen2009} suggested a degree discount heuristics to significantly decrease the running time by only considering the direct influence of a node to its one-hop neighbors, however, this method is tailored to influence maximization on uniform independent cascade model. Wang et al.~\cite{Wang2010} divided a network into communities and conducted Monte Carlo simulations within each community instead of the whole network. Luo et al.~\cite{Luo2012} conducted the greedy algorithm on a small set of nodes, consisting of the top nodes ranked by PageRank algorithm on social network. Kimura and Saito~\cite{Kimura2010} proposed the shortest-path based influence cascade models and provided efficient algorithms to compute the influence spread under these models. Instead of using the simple shortest path, PMIA algorithm~\cite{Chen2010,Wang2012} employed maximum influence paths for influence spread estimation, and this algorithm is believed to be the best heuristic algorithm so far. However, these heuristics may violate the guaranteed accuracy of greedy algorithm and thus one may concern about the reliability of these heuristics.


In addition, several influence maximization algorithms are beyond the framework of greedy algorithm. Jiang et al.~\cite{Jiang2011} suggested a simulated annealing approach with several heuristics to speed up influence spread estimation. Narayanam et al.~\cite{Narayanam2011} gave a way to improve the scalability of influence maximization using the concept of Shapley value borrowed from the cooperative game theory. Mathioudakis et al.~\cite{Mathioudakis2011} suggested removing some unimportant edges to accelerate influence computation algorithms.

Moreover, recently several works studied influence maximization problem in competitive environment~\cite{Bharathi2007,Carnes2007,Shirazipourazad2012}. Bharathi et al. modified the independent cascade model to the case of multiple competing innovations~\cite{Bharathi2007}. Carnes et al.~\cite{Carnes2007} extended the independent cascade model to a distance-based model and a wave propagation model. The two models are further studied by Shirazipourazad et al~\cite{Shirazipourazad2012}, and they tried to minimize the cost (the number of seed nodes selected) under a given competition goal. Since the object functions of those above extended independent cascade model still maintain the submodularity and monotonicity properties, greedy algorithms are used to achieve a guaranteed accuracy. In addition, some researchers studied influence spread limiting problem~\cite{Budak2011,He2012,Tripathy2010} under variants of the independent cascade model, but the submodularity and monotonicity properties of the object functions become difficult to be ensured.


\section{Static Greedy Algorithm}

\subsection{Influence maximization problem}

We consider the influence maximization problem with respect to the independent cascade model. For a directed graph $G = (V, E)$, each edge $\langle u,v\rangle \in E$ is associated with a probability $p(u,v)$. When $u$ is activated, it has one chance to activate $v$ with the successful rate $p(u,v)$, if $v$ has not been activated yet. The activation is fully determined by $p(u,v)$. Given a seed set $S$, its influence spread $I(S)$ is defined as the expected number of nodes eventually activated. The influence maximization problem aims at finding a set $S$ that maximizes $I(S)$, under the constraint that the size of $S$ is no larger than a predefined positive integer $k$.

To resolve the influence maximization problem, one needs to estimate $I(S)$ for any given $S$. However, it is intractable to exactly compute $I(S)$ on a typically sized graph. In practice, Monte Carlo methods are employed to estimate $I(S)$, and can be implemented in two different ways as follows:
\begin{itemize}

\item{Simulation.}
  The influence spread is obtained by directly simulating the random process of diffusion triggered by a given seed set $S$. Let $A_i$ denote the set of nodes newly activated in the $i$-th iteration and we have $A_0=S$. In the $(i+1)$-th iteration, a node $u \in A_i$ attempts to activate each inactive neighbor $v$ with the probability $p(u, v)$. If it succeeds, $v$ is added into $A_{i+1}$. The process is repeated until no activation is possible, and the number of eventually activated nodes is the influence spread of this single simulation. We run such simulations for many times and finally estimate the influence spread $I(S)$ by averaging over all simulations.

\item{Snapshot.}
  According to the characteristic of the IC model, whether $u$ successfully activates $v$ depends only on $p(u,v)$, like flipping a coin of bias $p(u,v)$. We can flip all coins a priori to produce a \emph{snapshot} $G' = (V, E')$, which is a subgraph of $G$ where an edge $\langle u,v\rangle$ is remained with the probability $p(u,v)$, and deleted otherwise. Such a snapshot provides an easy way to estimate the influence spread of $S$, which exactly equals to the number of nodes reachable from $S$. We produce plenty of snapshots and finally estimate the influence spread $I(S)$ by averaging over all snapshots.

\end{itemize}

Those two methods are essentially equivalent and either has its own advantage and disadvantage. For estimating the influence spread of a given seed set, the simulation method is faster, because it only needs to examine a small portion of edges while the snapshot method has to examine all the edges. For estimating the influence spreads of different seed sets, the snapshot method outperforms the simulation method in terms of time complexity, since each snapshot serves all seed sets.

\subsection{The submodularity property: the key to solve the scalability-accuracy dilemma}

For any greedy algorithm of influence maximization, it is required that the influence spread function $I(\cdot)$ is monotone and submodular to achieve a $(1-1/\mathrm{e})$-approximation~\cite{Nemhauser1978}. We say that a function $f(\cdot)$ is monotone if $f(S \cup \{v\}) \geq f(S)$ for any set $S$ and any element $v\notin S$, and $f(\cdot)$ is submodular if $f(S \cup \{v\})-f(S) \geq f(T \cup \{v\})-f(T)$ when $S \subseteq T$. The submodularity property is also explained as a natural ``diminishing return'' property. It has been proven that $I(\cdot)$ is monotone and submodular when its value can be \emph{exactly} estimated~\cite{Kempe2003,Mossel2007}. Unfortunately, things become different when Monte Carlo simulation is employed to \emph{approximately} estimate $I(\cdot)$.

\begin{figure}
\centering
\subfigure[Original graph]
{\label{fig:example:originalgraph} 
\includegraphics[width=0.31 \linewidth]{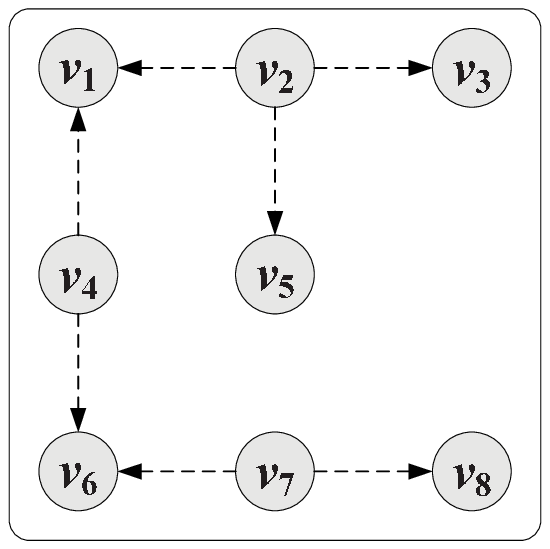}}
\label{fig:side:a}
\subfigure[Snapshot1]
{\label{fig:example:snapshot1} 
\includegraphics[width=0.31 \linewidth]{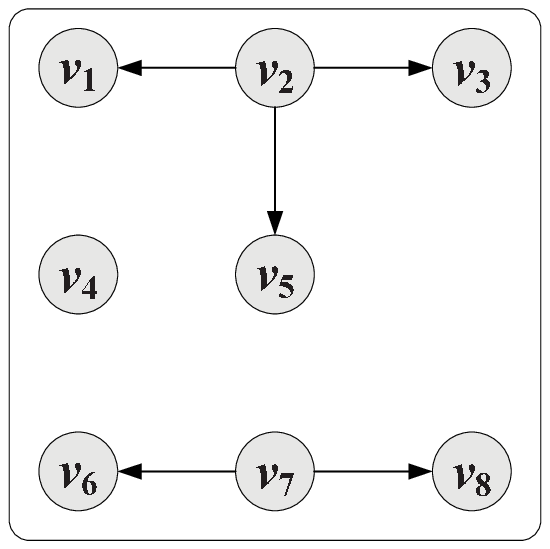}}
\subfigure[Snapshot2]
{\label{fig:example:snapshot2} 
\includegraphics[width=0.31 \linewidth]{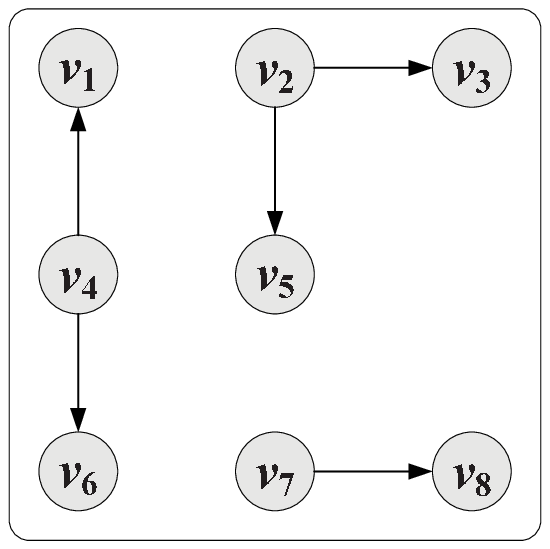}}
\caption{Illustrations of unguaranteed submodularity property.}\label{fig:example}
\end{figure}

Let us take a closer look. In existing greedy algorithms, different Monte Carlo simulations are conducted independently across different iterations. The spread along an edge $\langle u,v\rangle$ may fail in one Monte Carlo simulation and succeed in another Monte Carlo simulation. As a result, the marginal gain from adding $v$ to the seed set in the $i$-th iteration might be lower than the marginal gain from adding $v$ in the $(i+1)$-th iteration, i.e., $I(S_i \cup \{v\})-I(S_i) < I(S_{i+1} \cup \{v\})-I(S_{i+1})$ with $S_i \subset S_{i+1}$, which violates the submodularity property. For example, we produce Monte Carlo snapshots for a graph where each edge $\langle u,v\rangle$ associated with a uniform propagation probability $p(u, v)=0.5$ as shown in Figure~\ref{fig:example:originalgraph}. In the first iteration we produce one snapshot shown in Figure~\ref{fig:example:snapshot1}, and in the second iteration we produce another snapshot shown in Figure~\ref{fig:example:snapshot2}. We start from an empty seed set $S_0 = \emptyset$. Obviously $S_1 = \{v_2\}$, since $v_2$ has the largest influence spread in Figure~\ref{fig:example:snapshot1}. Now we check the marginal gains from adding $v_4$ in the two iterations, which are estimated on the two snapshots respectively.

\begin{eqnarray}
&&I(S_0 \cup \{v_4\}) - I(S_0) = I(\{v_4\}) - I(\emptyset)=1,\nonumber \\
&&I(S_1 \cup \{v_4\}) - I(S_1) = I(\{v_2,v_4\}) - I(\{v_2\})=3. \nonumber
\end{eqnarray}

The marginal gain from adding $v_4$ increases from $1$ to $3$, dissatisfying the submodularity requirement. The reason is that the estimation of influence spread of $v_4$, as well as that of $v_2$, differs between the two iterations with different snapshots being used. To summarize, producing different Monte Carlo simulations across different iterations brings the risk of unguaranteed submodularity. Similarly, the monotonicity property is also unguaranteed.

To reduce the pain from unguaranteed submodularity and monotonicity, one has to estimate the influence spread function $I(\cdot)$ exactly. For this purpose, existing greedy algorithms conduct an extremely large number (typically $10,000$ or $20,000$) of Monte Carlo simulations in every iteration. However, the submodularity and monotonicity properties can only be guaranteed with a certain probability in this way because of the finite number of Monte Carlo simulations. As a result, to achieve the guaranteed $(1-1/\mathrm{e})$-approximation, existing greedy algorithms have to bear the expensive computational cost for conducting huge number of Monte Carlo simulations. This poses the scalability-accuracy dilemma suffered by existing greedy algorithms. 

\subsection{Description of static greedy algorithm}

We have pointed out that the key for combating the scalability-accuracy dilemma is ensuring that the estimated influence spreads of any seed set are the same in different iterations, so as to overcome the unguaranteed submodularity and monotonicity rooted in different Monte Carlo simulations conducted in different iterations of greedy algorithms. We propose a solution to guarantee submodularity and monotoncity in a simpler but more effective way. Instead of producing a huge number of Monte Carlo simulations in every iterations, we produce a (not very large) number of Monte Carlo snapshots at the very beginning, and use the same set of snapshots in all iterations. Those snapshots are called ``static'', results in the \textbf{StaticGreedy} algorithm. That algorithm ensures that the estimated influence spreads of any seed set are exactly the same in different iterations, and thus guarantees submodularity and monotonicity properties. Avoiding a huge number of Monte Carlo simulations needed in every iterations, our algorithm brings the possibility to significantly reduce the computational expense without loss of accuracy.

Given an underlying social network $G$ and a positive integer $k$, the StaticGreedy algorithm runs in the following two stages to seek for a seed set $S$ that maximizes the influence spread $I(S)$:

\begin{enumerate}
\item Static snapshots: Select a value of $R$, the number of Monte Carlo snapshots, then randomly sample $R$ snapshots from the underlying social network $G$. For each snapshot, each edge $\langle u,v\rangle$ is sampled according to its associated probability $p(u,v)$;
\item Greedy selection: Start from an empty seed set $S$, iteratively add one node a time into $S$ such that the newly added node provides the largest marginal gain of $I(S)$, which is estimated on the $R$ snapshots. This process continues until $k$ seed nodes are selected.
\end{enumerate}

The StaticGreedy algorithm is formally described in Algorithm~\ref{algorithm:staticgreedy}. Two main differences between this algorithm and existing greedy algorithms include: (1) Monte Carlo simulations are conducted in static snapshot manner, which are sampled before the greedy process of selecting seed nodes, as is shown in line \ref{algorithm:staticgreedy:sampling-start} to \ref{algorithm:staticgreedy:sampling-end}; (2) The same set of snapshots are reused in every iteration to estimate the influence spread $I(S)$, where explains the meaning of ``static''.

\begin{algorithm}[t]
\caption{StaticGreedy($G$,$k$,$R$)}\label{algorithm:staticgreedy}
\begin{algorithmic}[1]
    \STATE initialize $S = \varnothing$
    \FOR{$i = 1$ to $R$} \label{algorithm:staticgreedy:sampling-start}
        \STATE generate snapshot $G'_i$ by removing each edge $\langle u,v\rangle$ from $G$ with probability $1-p(u, v)$
    \ENDFOR \label{algorithm:staticgreedy:sampling-end}
    \FOR{$i = 1$ to $k$}
        \STATE set $s_v$$= 0$ for all $v \in {V \setminus S}$       //$s_v$ stores the influence spread after adding node $v$
        \FOR{$j = 1$ to $R$}
           \FORALL {$v \in V \setminus S$}
                    \STATE $s_v$ += $|R(G'_j, S \cup \{v\})|$       //$R(G'_j,S \cup \{v\})$ is the influence spread of $S \cup \{v\}$ in snapshot $G'_j$
           \ENDFOR
        \ENDFOR
        \STATE $S=S \cup \{\arg\max\limits_{v \in {V \setminus S }}\{s_v/R\}$\}
    \ENDFOR
    \STATE output S
\end{algorithmic}
\end{algorithm}

Both our StaticGreedy algorithm and conventional greedy algorithms provide a $(1-1/\mathrm{e}-\epsilon)$-approximation to the optimal solution of influence maximization. The main difference lies in the origin of $\epsilon$. For conventional greedy algorithms, $\epsilon$ depends on the extent to which the submodularity is guaranteed and it generally requires a huge number of Monte Carlo simulations, typically in the magnitude of 10,000. For StaticGreedy algorithm, $\epsilon$ is caused by the variance of the unbiased estimation to the optimal influence spread using finite static snapshots. In practice, a small $\epsilon$ is usually achieved using a small number of static snapshots, e.g., 100. In this way, StaticGreedy algorithm efficiently solves the scalability-accuracy dilemma suffered by conventional greedy algorithms for influence maximization.

\subsection{Analysis of the StaticGreedy algorithm}

\subsubsection{Accuracy}
To clarify the effectiveness of the StaticGreedy algorithm compared with traditional greedy algorithms, we illustrate the accuracy of these algorithms with respect to the number $R$ of Monte Carlo simulations on a benchmark network NetHEPT. This network consists of tens of thousands of physics researchers and their co-authorship relations. The baseline greedy algorithm is the CELFGreedy, which is the general greedy algorithm with CELF optimization, and the NewGreedy, which is a snapshot-based greedy algorithm reusing snapshots for influence spread estimation within the same iteration. We employ the NewGreedy algorithm as a comparison to show that our StaticGreedy is fundamentally different from existing snapshot-based greedy algorithm. The comparisons are conducted with respect to two commonly-used IC models: the uniform independent cascade (UIC) model with $p=0.01$ and the weighted independent cascade (WIC) model~\cite{Kempe2003} with $p(u,v)=1/d_v$, where $d_v$ is the indegree of node $v$.

Since the optimal influence spread is unknown to us, the ground truth we use here is the influence spread of the solution $S_k^*$ with the set size $k$, obtained by the CELFGreedy algorithm with typical setting, i.e., $R$ = 20,000. To evaluate the relative difference between the influence spread $I(S_{R,k})$ and the ground truth, we use a measure $d_{R,k}$ defined as
\begin{equation*}\label{relativedifference}
    d_{R,k} = \frac{I(S_k^*)-I(S_{R,k})}{I(S_k^*)},
\end{equation*}
where $S_{R,k}$ is the set of seed nodes obtained by a greedy algorithm with a given $R$, and $k$ is the size of seed set. For a given $R$, we run each of the three greedy algorithms for $100$ times to calculate the average relative difference. Here, we only report the results with $k=50$ since the results for other $k$ are similar.

\begin{figure}[t]
\centering
{\includegraphics[width=0.8 \linewidth]{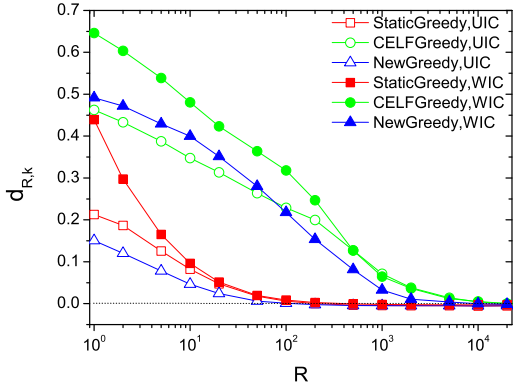}}
\caption{The relationship between $d_{R,k}$ and $R$ on NetHEPT network. } \label{fig:ThreeGreedyRD}
\end{figure}

As shown in Figure~\ref{fig:ThreeGreedyRD}, for both the UIC model and the WIC model, the StaticGreedy algorithm quickly approaches to the ground truth while the CELFGreedy algorithm converges slowly. This confirms that the StaticGreedy algorithm can achieve good accuracy even when the number $R$ of Monte Carlo simulations is small,
e.g., $R = 100$. Moreover, the accuracy of StaticGreedy algorithm consistently outperforms the accuracy of CELFGreedy algorithm. The NewGreedy algorithm performs very differently for the UIC model and the WIC model. It needs a large $R$ for the WIC model although it works well for the UIC model. Furthermore, the smaller value of $R$ does not indicate that the NewGreedy algorithm is more effective than the StaticGreedy algorithm, because the NewGreedy algorithm needs $k*R$ Monte Carlo simulations with each iteration using $R$ simulations. We will give more discussions about this point later. As a conclusion, only the StaticGreedy algorithm exhibits consistently good performance for the two models.

We further evaluate the accuracy of the StaticGreedy algorithm with respect to the size $k$ of seed set. For this purpose, we define $R_{min}$ as the minimal $R$ satisfying $d_{R,k} \leq 0.005$. As shown in Figure~\ref{fig:ThreeGreedyRmin}, the values of $R_{min}$ for the StaticGreedy algorithm are consistently smaller than the values of $R_{min}$ for the CELFGreedy algorithm. The NewGreedy algorithm again performs differently on the two models.

\begin{figure}[t]
\centering
{\label{fig:ThreeGreedyRmin} 
\includegraphics[width=0.8 \linewidth]{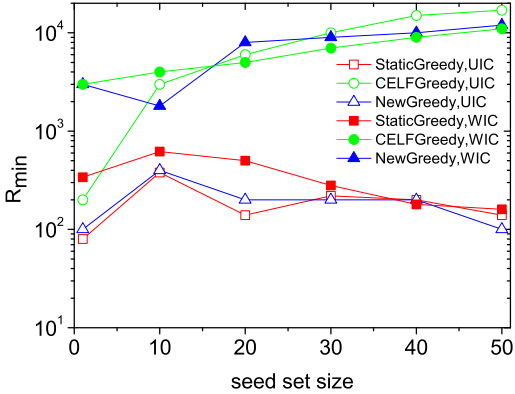}}

\caption{Minimal number of snapshots needed to accurately find a solution.} \label{fig:ThreeGreedyRmin} 
\end{figure}

\begin{figure}[t]
\centering
\includegraphics[width=0.75 \linewidth]{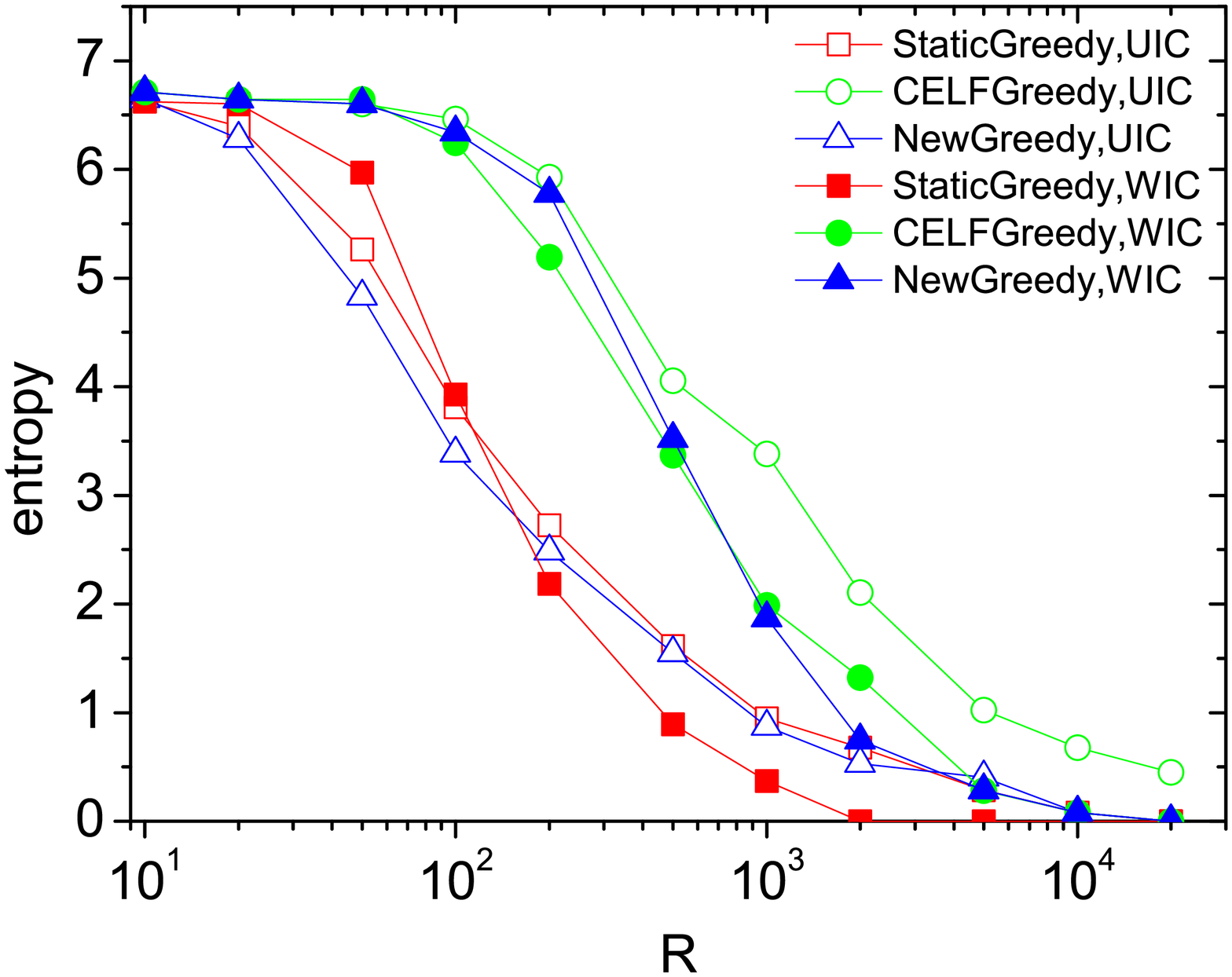}
\caption{The entropy of the solution space with respect to $R$.} \label{fig:solutionspace} 
\end{figure}

To further understand the finding that the StaticGreedy algorithm can achieve a rapid convergence with respect to $R$, we introduce a measure $H_{R,k}(S)$ to analyze the solution space of greedy algorithm, which is defined as
\begin{equation*}\label{solutionspaceentropy}
    H_{R,k}(S) = -\sum{p(S)\log{p(S)}},
\end{equation*}
where $p(S)$ is the fraction of a certain solution $S$ relative to the size of solution space according to the setting of a given $R$ and $k$ for a greedy algorithm. $H_{R,k}(S)$ is a kind of entropy, which characterizes the heterogeneity of a probability distribution. A large value of $H_{R,k}(S)$ means much uncertainty of the solution. When the value of $H_{R,k}(S)=0$, the algorithm converges to a unique solution. Actually, there are always many solutions with very close influence, and the number of different solutions is always larger when the network is larger or $k$ becomes larger. Here, we choose $k=5$ to illustrate the advantage of StaticGreedy algorithm over the other two greedy algorithms. For each $R$, we run the algorithm for $100$ times and calculate the $H_{R,k=5}(S)$ according to the obtained 100 solutions. As shown in Figure~\ref{fig:solutionspace}, the solution space of StaticGreedy algorithm narrows quickly, while the CELFGreedy shows a slow convergence. For the UIC model, the trend of $H(S)$ for the StaticGreedy algorithm and the NewGreedy algorithm is similar, explaining why the two algorithms need a similar $R_{min}$ under this model. For the WIC model, the $H(S)$ for the NewGreedy algorithm and the CELFGreedy algorithm are similar and narrow slowly, while $H(S)$ converges quickly for the StaticGreedy algorithm. In sum, with the strictly guaranteed submodularity property, the StaticGreedy algorithm can always achieve a rapid convergence of the solution space.

According to the above analysis, we can see that the StaticGreedy algorithm is essentially different from the NewGreedy algorithm. The NewGreedy algorithm aims to reduce the computational cost by simultaneously estimating the influence spread of many seed sets in the same iteration, while the submodularity property is not maintained since different iterations do not share the results of  Monte Carlo simulations as done by the StaticGreedy algorithm.

\subsubsection{Scalability}

Now we analyze the time complexity of the StaticGreedy algorithm. Since the number of Monte Carlo simulations for influence spread estimation, $R$, is significant different for our StaticGreedy algorithm and other greedy algorithms, for clarity, we use $R$ to denote the number of Monte Carlo simulations required by existing greedy algorithms and use $R'$ to denote the number of Monte Carlo simulations required by our StaticGreedy algorithm. In addition, $n$ is the number of nodes in the underlying influence network, $m$ is the number of edges in the network, $m'$ is the average number of active edges in the snapshots obtained by sampling the influence network, and $k$ is the number of seed nodes. For the StaticGreedy algorithm, the time complexity includes two parts: firstly, the time complexity of generating $R'$ snapshots is $O(R'm)$; secondly, it takes $O(knR'm')$ time to select seed nodes in greedy manner on those static snapshots. Thus, the total time complexity is $O(R'm+knR'm')$. For the space complexity of StaticGreedy algorithm is $O(R'm')$, which is used to store the $R'$ snapshots. The comparison with the general greedy algorithm~\cite{Kempe2003} and the NewGreedy algorithm~\cite{Chen2009} is given in Table~\ref{table:timespace}. 

\begin{figure}[t]
\centering
{\label{fig:ThreeGreedyRminrunningtime} 
\includegraphics[width=0.8 \linewidth]{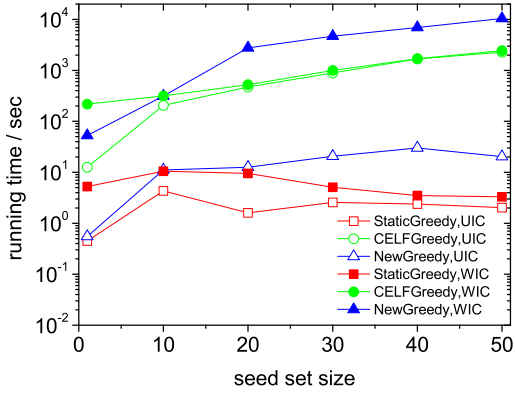}}
\caption{The running time of each greedy algorithm with their respective $R_{min}$} \label{fig:ThreeGreedyRminRunningtime} 
\end{figure}

Figure~\ref{fig:ThreeGreedyRminRunningtime} shows the running time of each greedy algorithm with their respective $R_{min}$ for different $k$. The StaticGreedy algorithm outperforms the other two greedy algorithms, and runs much faster than the CELFGreedy algorithm. Although the NewGreedy algorithm has a similar small $R_{min}$ to the StaticGreedy algorithm, its time-consuming is still larger than the StaticGreedy algorithm, which is mainly because that the NewGreedy algorithm needs to do $R$ Monte Carlo simulations in every iteration, while StaticGreedy algorithm only need to do $R$ times at the very beginning. Moreover, we later propose an improved version of the StaticGreedy algorithm, which can further effectively decrease the running time of the current StaticGreedy algorithm.

The time complexity of the StaticGreedy algorithm can be further reduced by employing the CELF optimization and other optimization strategies. In the next section, we give a dynamic update strategy to improve the efficiency of the StaticGreedy algorithm.

\begin{table}
\centering \caption{Time and space complexity of algorithms} \label{table:timespace}
\begin{tabular}{p{2.2cm}cc}
\toprule
Algorithms                   & Time complexity                          & Space complexity   \\
\midrule
\textbf{StaticGreedy}        & $O(R'm+knR'm')$                         & $O(R'm')$      \\
\textbf{GeneralGreedy}       & $O(knRm)$                                & $O(m)$      \\
\textbf{NewGreedy}        & $O(kRTm)$                         & $O(m)$      \\
\bottomrule
\end{tabular}
\end{table}



\subsubsection{Discussions}

One may ask the question: why the StaticGreedy algorithm can achieve the high accuracy with a small $R$? Indeed, a small $R$ may result in the inaccurate estimation of the influence spread of a given seed set. However, as we show here, the inaccurate estimation matters little at finding the solution of influence maximization.  Basically, the reason lies in that the influence maximization aims to find a set of nodes rather than a ranked set of nodes. The inaccurate estimation of influence spread may alter the order of nodes in the seed set while has little influence on the set of nodes.

The idea behind the StaticGreedy method for independent cascade model can be easily generalized to the linear threshold model. In this paper, the details are omitted with the limitation of space.



In addition, it is unclear on how to determine a suitable $R$ at present. How do we determine the minimum $R$ for a specific network and a given spread model? What are the factors affecting $R$ in StaticGreedy or previous greedy algorithms? We leave these interesting questions as open problems in the future.

\section{Speeding up the StaticGreedy}

In this section, we propose a dynamic update strategy to speed up the proposed static greedy algorithm. This strategy exploits the advantage of static snapshots and calculates the marginal gain in an efficient incremental manner. Specifically, when a node $v^*$ is selected as a seed node, we directly discount the marginal gain of other nodes by the marginal gain shared by these nodes and $v^*$.

For a snapshot $G'_i$, we use $R(G'_i, v)$ to denote the set of nodes which are reachable from $v$ and use $U(G'_i, v)$ to denote the set of nodes from which $v$ can be reached. In the first iteration, the marginal gain of $v$ is $|R(G'_i, v)|$. In our dynamic update strategy, when $v^*$ is selected as a seed node, we find the set $U(G'_i, w)$ for each node $w \in R(G'_i, v^*)$. Then, for every $u \in U(G'_i, w)$, we delete $w$ from $R(G'_i, u)$. The size of the remained $R(G'_i,u)$ reflects the marginal gain of $u$ in the next iteration. In this way, we can maintain a dynamically updated marginal gain for each node to avoid calculating the marginal gain from scratch. The detailed implementation of the improved static algorithm, namely \textbf{StaticGreedyDU}, is given in Algorithm~\ref{algorithm:static.snapshot.dynamicupdate}.

Now we analyze the time and space complexity of the StaticGreedyDU algorithm. For undirected graphs, $R(G'_i, v)$ is the same to $U(G'_i, v)$. We only need to store the information of connected components for each snapshot. Thus, the space complexity is $O(R'n)$. The time complexity is $O(R'm)$ for generating $R'$ snapshots and calculating the initial marginal gain. The time complexity is $O(kn)$ for updating the marginal gain of all the related nodes. Thus, the total time complexity is $O(R'm+kn)$. For directed graphs, let $n_T = \max_{v \in V}R(G'_i,v)$, $n_U = \max_{v \in V}U(G'_i,v)$. Since it needs to store $R(G'_i, v)$ and $U(G'_i, v)$ for each node, the space complexity is $O(R'nn_{T}+R'nn_{U})$. Assume the maximum running time to compute $R(G'_i, v)$ and $U(G'_i, v)$ is $t_{T}$ and $t_{U}$ respectively. The time complexity is $O(R'm)$ for generating snapshots, $O(R'nt_{T}+R'nt_{U})$ for computing the initial incremental influence spread, and $O(kR'n_{T}n_{U})$ for updating the marginal gains. Thus, the total time complexity is $O(R'm+R'nt_{T}+R'nt_{U}+kR'n_{T}n_{U})$ for directed graphs. Note that $n_T$, $n_U$, $t_T$ and $t_U$ are usually very small in real world networks since these networks are usually sparse.

\begin{algorithm}[t]
\caption{StaticGreedyDU($G$,$k$,$R$)}\label{algorithm:static.snapshot.dynamicupdate}
\begin{algorithmic}[1]
    \STATE initialize $S = \varnothing$
    \STATE set the marginal gain $s_v = 0$ for all $v \in V$
    \FOR{$i = 1$ to $R$}
        \STATE generate $G'_i$
        \STATE compute and record $R(G'_i, v)$ and $U(G'_i, v)$ for all $v \in V$
        \FOR{each node $v \in V$}
           \STATE $s_v$ += $|R(G'_i,v)|$
        \ENDFOR
    \ENDFOR
    \FOR{$r = 1$ to $k$}

        \STATE $v^* =\arg\max\limits_{v \in {V \setminus S }}\{s_v\}$\
        \STATE $S = S \cup \{v^*\}$
        \FOR{$i = 1$ to $R$}
           \FOR{each node $w \in R(G'_i, v^*)$}
              \FOR{each node $u \in U(G'_i, w)$}
               \STATE delete $w$ from $R(G'_i, u)$
               \STATE $s_u$ = $s_u$-1
              \ENDFOR
           \ENDFOR
        \ENDFOR
    \ENDFOR
    \STATE output S.
\end{algorithmic}
\end{algorithm}

\section{Experiment}
In this section, we conduct experiments on several real-world networks to compare our StaticGreedy algorithm with a number of existing algorithms. The experiments aim at illustrating the performance of our algorithm comparing to other algorithms from the following two aspects: (a) accuracy at finding the seed nodes maximizing the influence spread, (b) scalability.

\subsection{Experiment setup}

\noindent \textbf{Datasets.} Six real world networks are employed to demonstrate the performance of our algorithms by comparing with other existing algorithms. These networks include three undirected scientific collaboration networks and three directed online social networks. In the three scientific collaboration networks, namely NetHEPT, NetPHY, and DBLP~\footnote{The three scientific collaboration networks are downloaded from http://research. microsoft.com/en-us/people/weic/. Those networks are actually multigraphs, where parallel edges between two nodes denoting the number of papers coauthored by the two authors. We view parallel edges between two nodes as a single edge to simplify.}, nodes represent authors and edges represent coauthor relationships among authors. All of those $6$ networks are undirected. NetHEPT is extracted from the ``High Energy Physics - Theory'' section of the e-print arXiv website~\footnote{http://www.arXiv.org} between 1991 and 2003. NetPHY is constructed from the full paper list of the ``Physics'' section of the arXiv website. DBLP, much larger than the former two scientific collaboration networks, is extracted from the DBLP Computer Science Bibliography~\footnote{http://www.informatik.uni-trier.de/~ley/db/}.
The three online social networks Epinions, Slashdot, and Douban~\footnote{Epinions and Slashdot can be downloaded from http://snap.stanford.edu/data/. The last one can be obtained on demand via email to the authors.} are collected from the websites Epinions.com, Slashdot.com, and Douban.com. In the Epinions network, an edge $\langle u,v\rangle$ means that a user $u$ trusts another user $v$. Slashdot is a friend network extracted from a technology-related news website Slashdot.com. In the Douban network~\cite{Huang2012} an edge $\langle u,v\rangle$ means that a user $u$ follows another user $v$. All the three online social networks are directed. Those $6$ networks are representative networks, covering a variety of networks with different kinds of relations and different sizes ranging from tens of thousands of edges to millions. Basic statistics of those networks are given in Table~\ref{table:statisticsoftestnetworks}.

\begin{table}
\centering \caption{Statistics of six test real world networks.} \label{table:statisticsoftestnetworks}
\begin{tabular}{p{1.5cm}rrc}
\toprule
Datasets         & \#Nodes   & \#Edges   & Directed? \\
\midrule
NetHEPT         & 15K      & 59K       &undirected \\
NetPHY          & 37K      & 231K      &undirected \\
DBLP            & 655K     & 2M        &undirected \\
Epinions        & 76K      & 509K      & directed  \\
Slashdot        & 77K      & 905K      & directed  \\
Douban          & 552K     & 22M       & directed  \\
\bottomrule
\end{tabular}
\end{table}

\begin{figure*}[!tbh]
\centering
\subfigure[NetHEPT]
{\label{fig:performanceundirected:NetHEPT} 
\includegraphics[width=0.31 \linewidth]{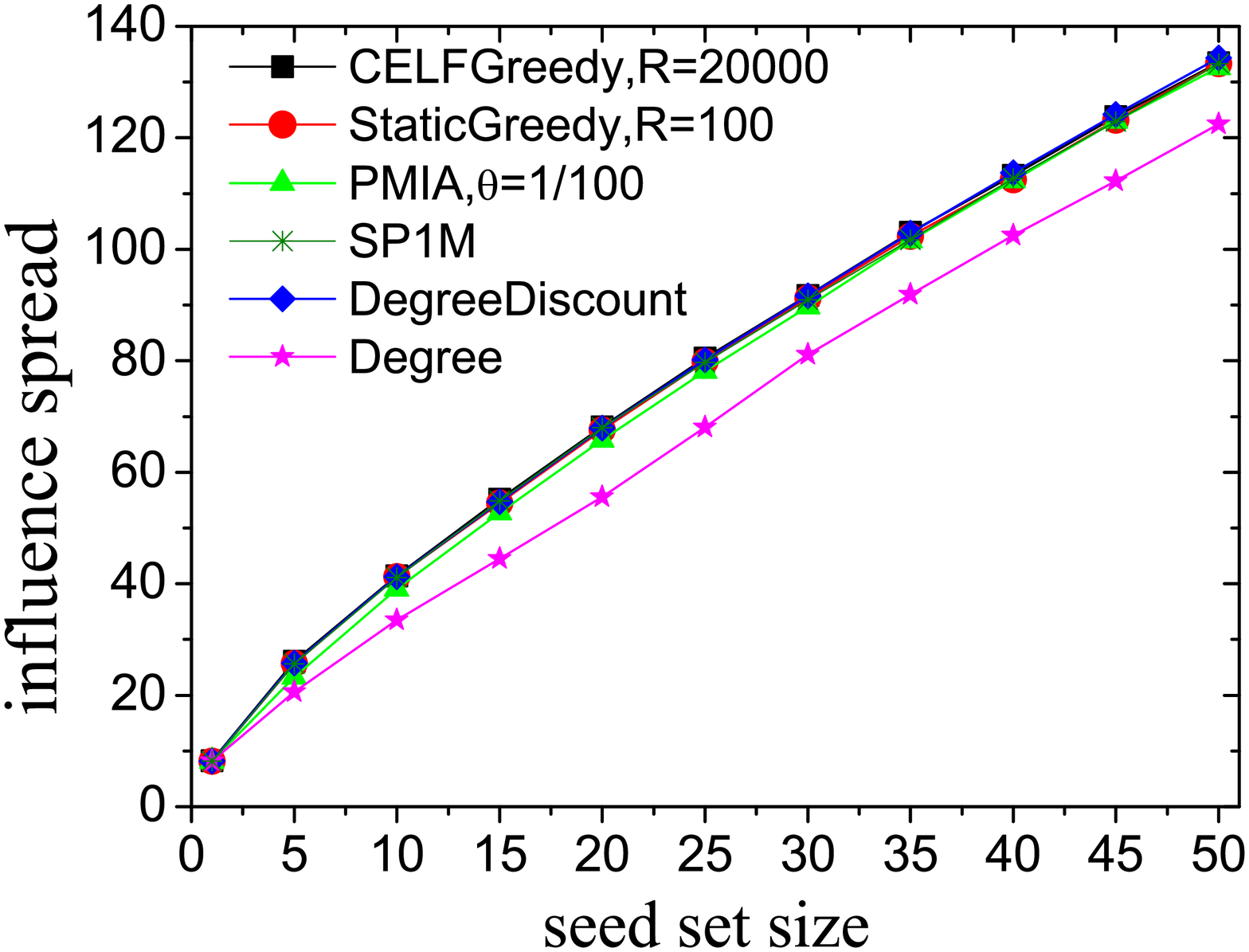}}
\subfigure[NetPHY]
{\label{fig:performanceundirected:NetPHY} 
\includegraphics[width=0.31 \linewidth]{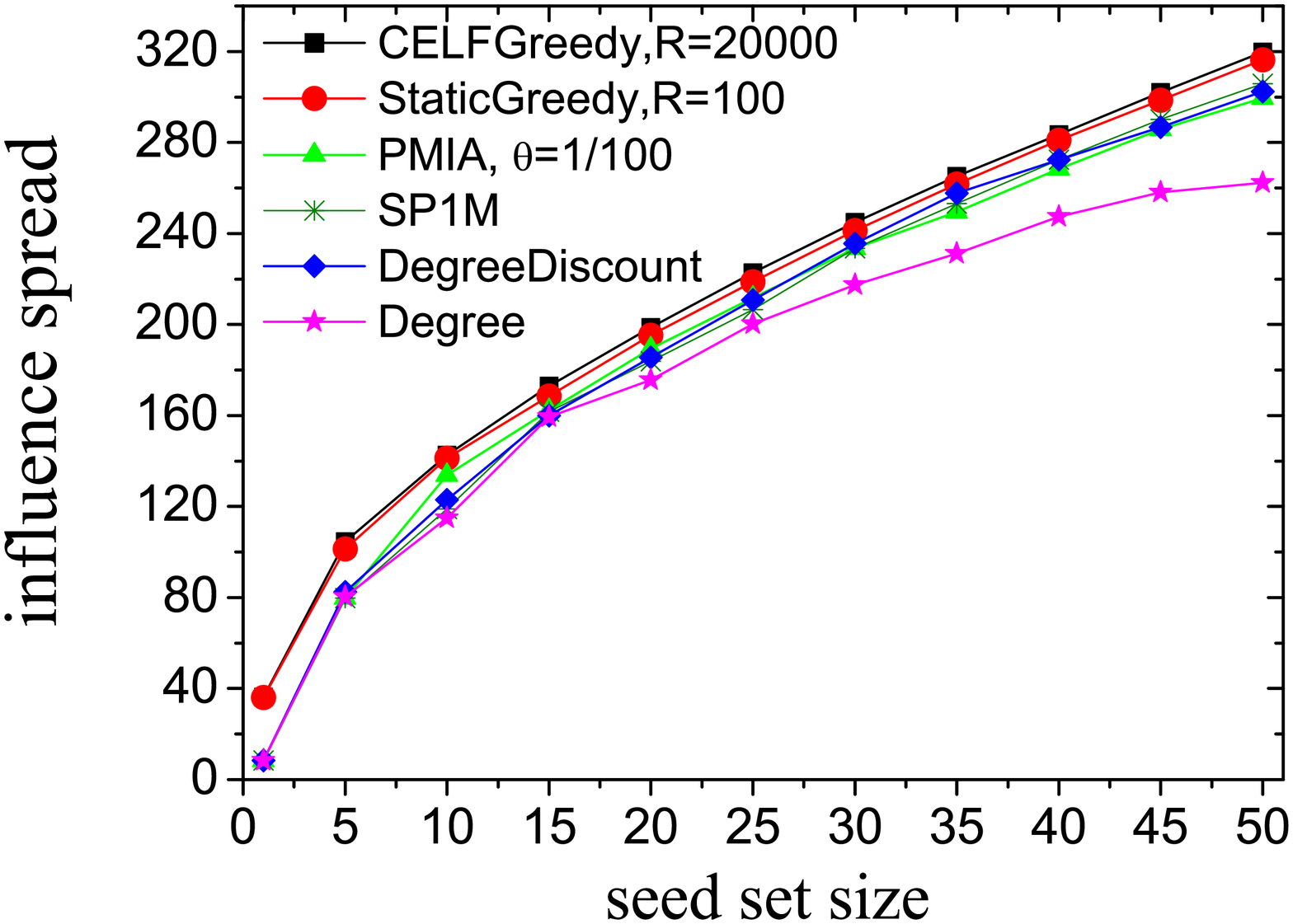}}
\label{fig:performanceundirected.uic:b}
\subfigure[DBLP]
{\label{fig:performanceundirected:DBLP} 
\includegraphics[width=0.31 \linewidth]{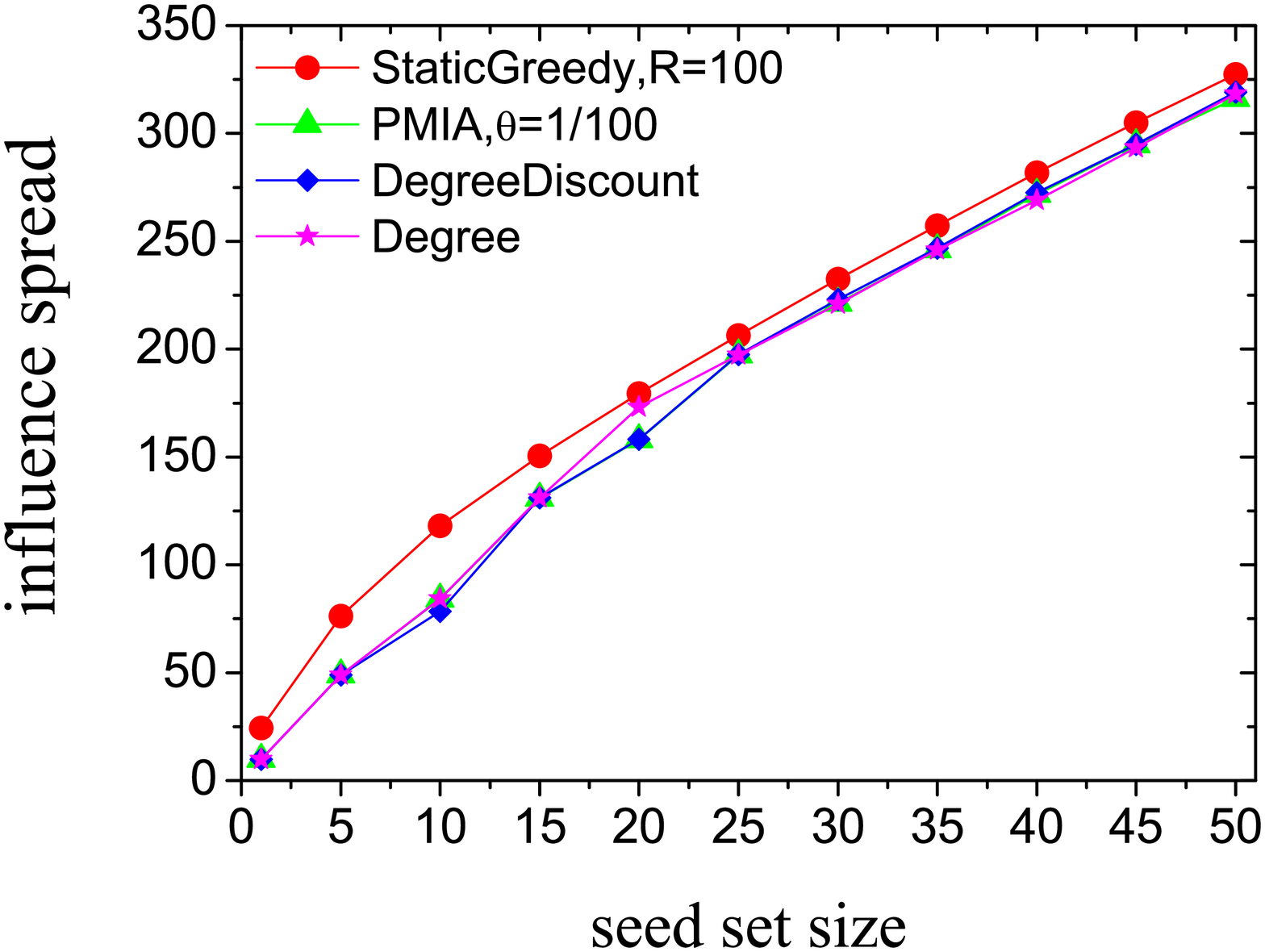}}
\label{fig:performanceundirected.uic:c}
\subfigure[Epinions]
{\label{fig:subfig:NetHEPT} 
\includegraphics[width=0.31 \linewidth]{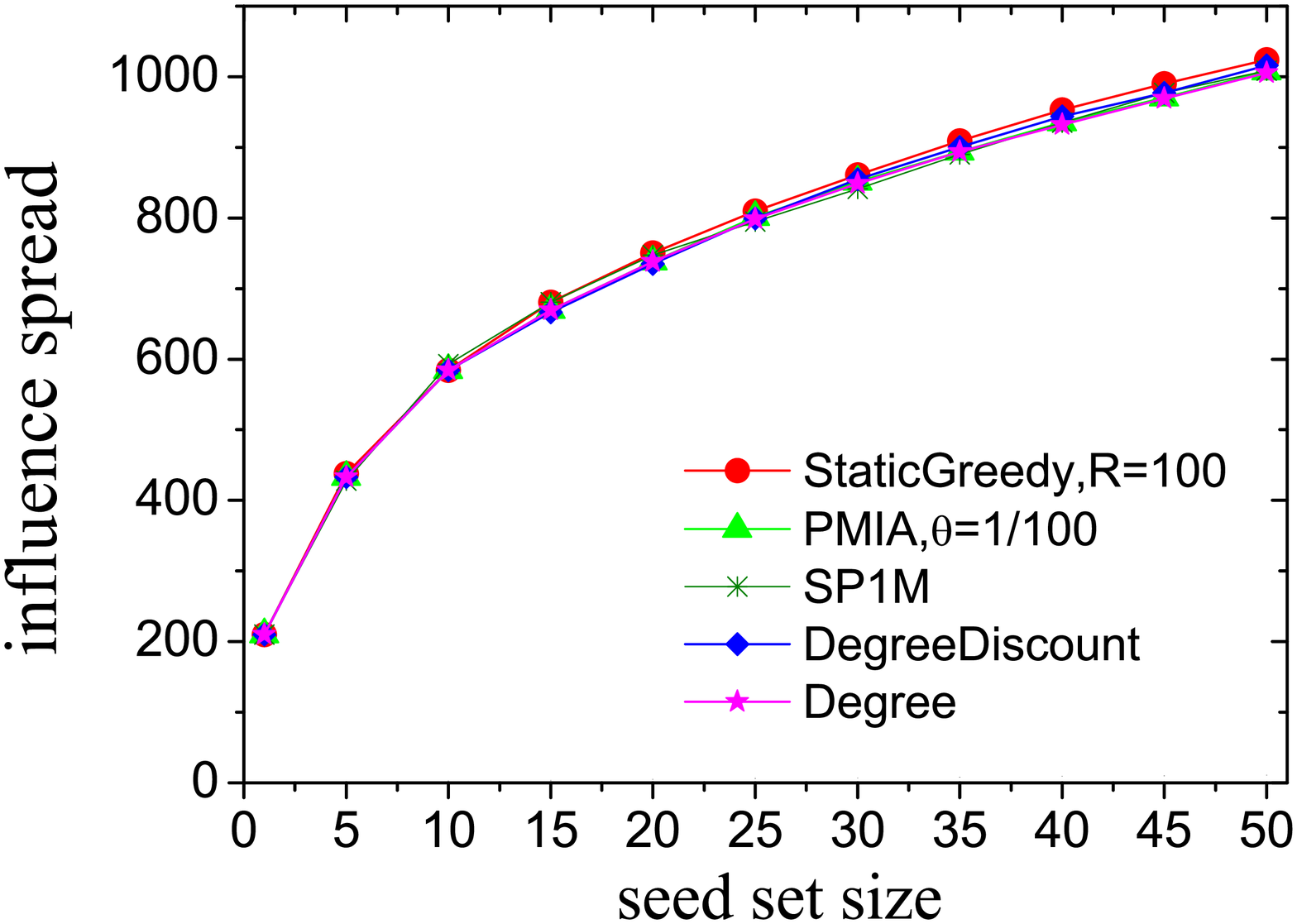}}
\label{fig:side:a}
\subfigure[Slashdot]
{\label{fig:subfig:NetPHY} 
\includegraphics[width=0.31 \linewidth]{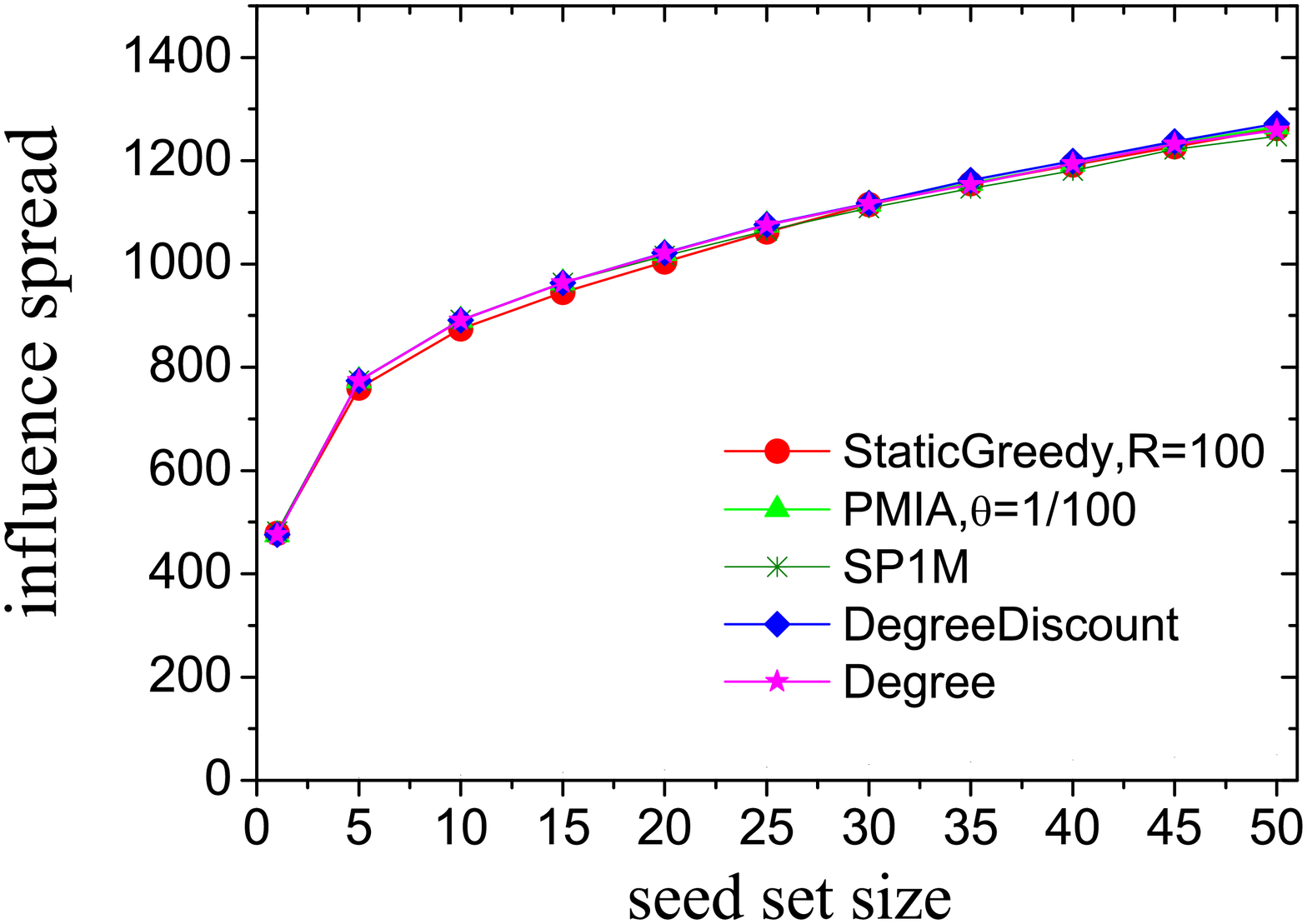}}
\subfigure[Douban]
{\label{fig:subfig:DBLP} 
\includegraphics[width=0.31 \linewidth]{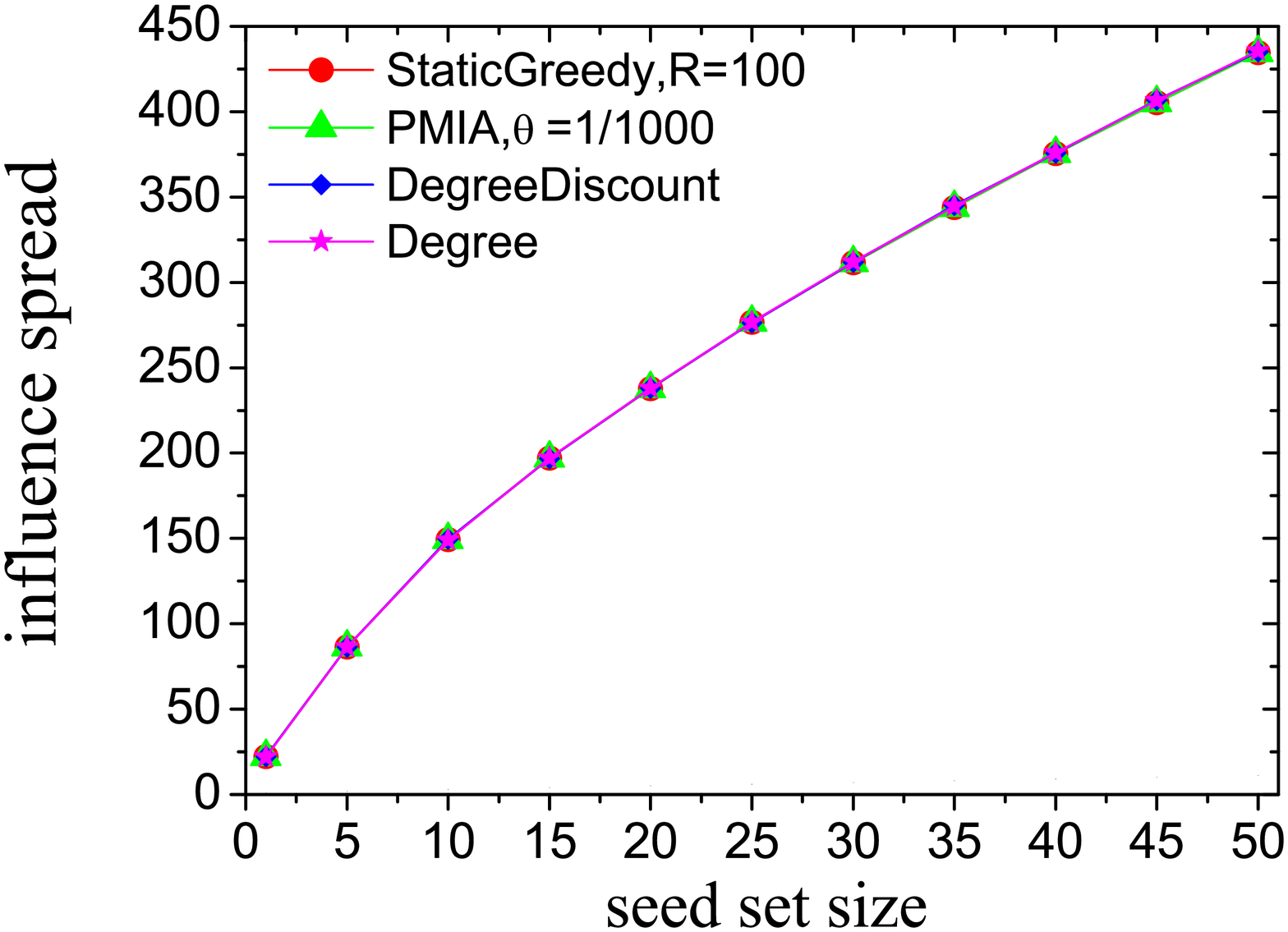}}
\caption{Influence spread under UIC model on six datasets.} \label{fig:performanceundirected.uic} 
\end{figure*}

\noindent \textbf{Influence spread models.} The algorithms are evaluated with two commonly used implementations of independent cascade model: the uniform independent cascade model (UIC) and the weighted independent cascade model (WIC). With UIC, the propagation probability on every edge is assigned with a uniform value. We assign $p(\cdot,\cdot)=0.001$ for Douban and $p(\cdot,\cdot)=0.01$ for other networks, because the average degree of Douban is roughly ten times than that of any other network. With WIC, the propagation probability on every edge could be assigned with different values. We follow a typical configuration to assign $p(u,v)=1/d_v$, where $d_v$ is the indegree of node $v$.

\noindent \textbf{Algorithms.} A total of six algorithms are tested, including our algorithm, a greedy algorithm CELFGreedy, as well as four heuristic algorithms PMIA, SP1M, DegreeDiscount and Degree.

\begin{itemize}
\item \textbf{StaticGreedy}
The algorithm proposed in this paper. We set $R=100$, i.e., $100$ snapshots in the whole process for any network.
\item \textbf{CELFGreedy}
The greedy algorithm with the CELF optimization~\cite{Leskovec2007}. We set $R=20,000$ as its recommended value to obtain accurate estimation, i.e., $20,000$ simulations for each candidate node in each iteration.
\item \textbf{PMIA}
The heuristic employs maximum influence paths for influence spread estimation \cite{Chen2010,Wang2012}. We set the value of $\theta=1/1000$ for Douban and $\theta=1/100$ for other networks as suggested in ~\cite{Chen2010}.
\item \textbf{SP1M}
A shortest-path based heuristic enhanced with the lazy-forward optimization~\cite{Kimura2006}.
\item \textbf{DegreeDiscount}
The heuristic considers the direct influence of a node to its one-hop neighbors~\cite{Chen2009}.
\item \textbf{Degree}
The heuristic simply selects seed nodes according to the degree of nodes in undirected networks or the outdegree in directed networks.
\end{itemize}

We also test the NewGreedy algorithm and the MixedGreedy algorithm on these datasets, and the results are similar to the CELFGreedy algorithm, hence we omit the two greedy algorithms. Since the PMIA heuristic is the state-of-the-art heuristic~\cite{Chen2010}, we do not implement more heuristics such as distance centrality, betweenness centrality, or PageRank-based heuristics. All experiments are conducted on a server with 2.0GHz Quad-Core Intel Xeon X7550 and 64G memory.

\subsection{Experimental results}

We run tests on the six datasets and two IC models. The tested seed size $k$ are $1$, $5$, $10$, $...$, up to $50$. For the comparison of running time, we only consider the seed size $k=50$. 

\subsubsection{Accuracy comparison}
\begin{figure*}[!htb]
\centering
\subfigure[NetHEPT]
{\label{fig:performanceundirected.wic:NetHEPT} 
\includegraphics[width=0.31 \linewidth]{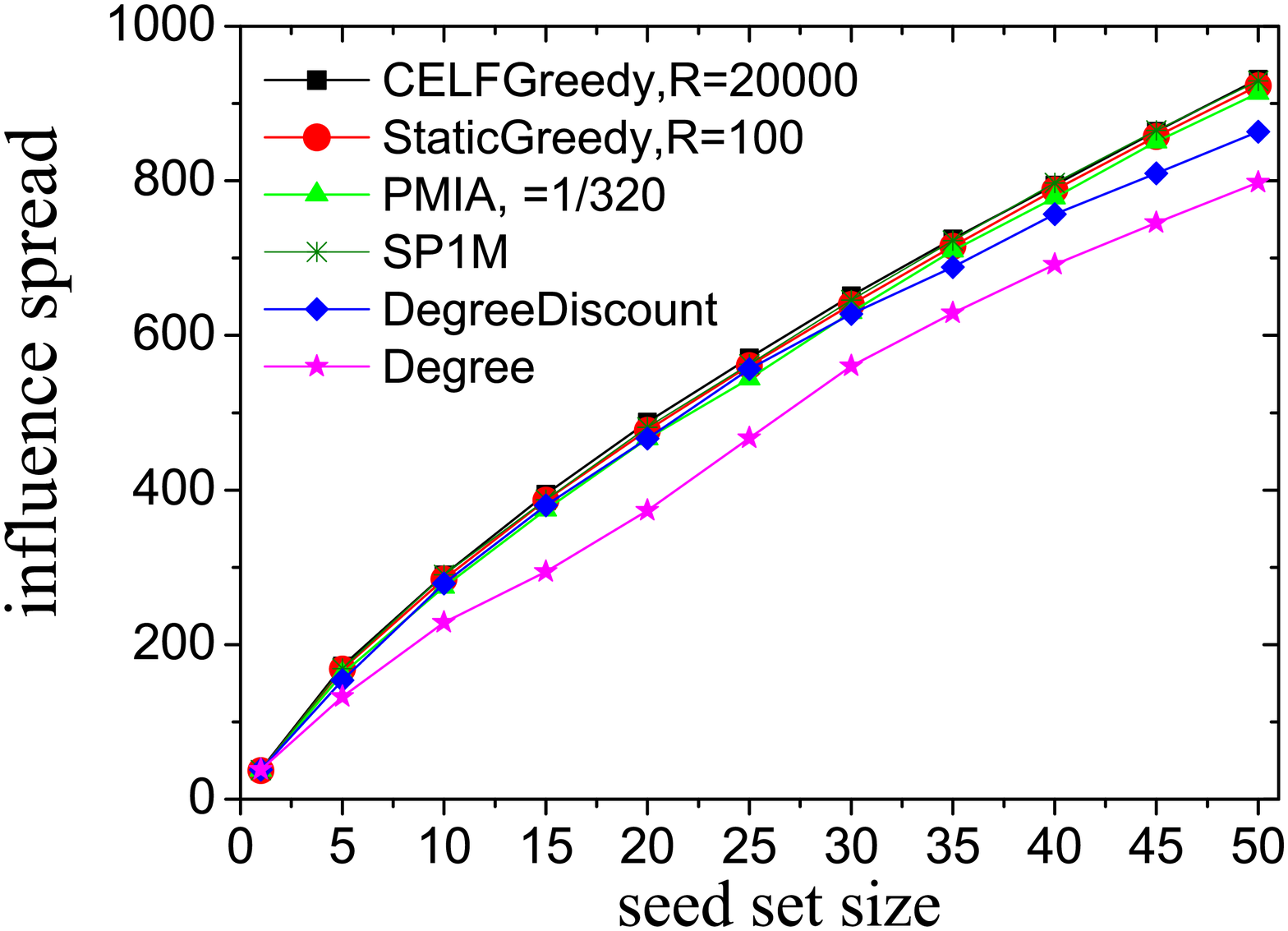}}
\label{fig:side:a}
\subfigure[NetPHY]
{\label{fig:performanceundirected.wic:NetPHY} 
\includegraphics[width=0.31 \linewidth]{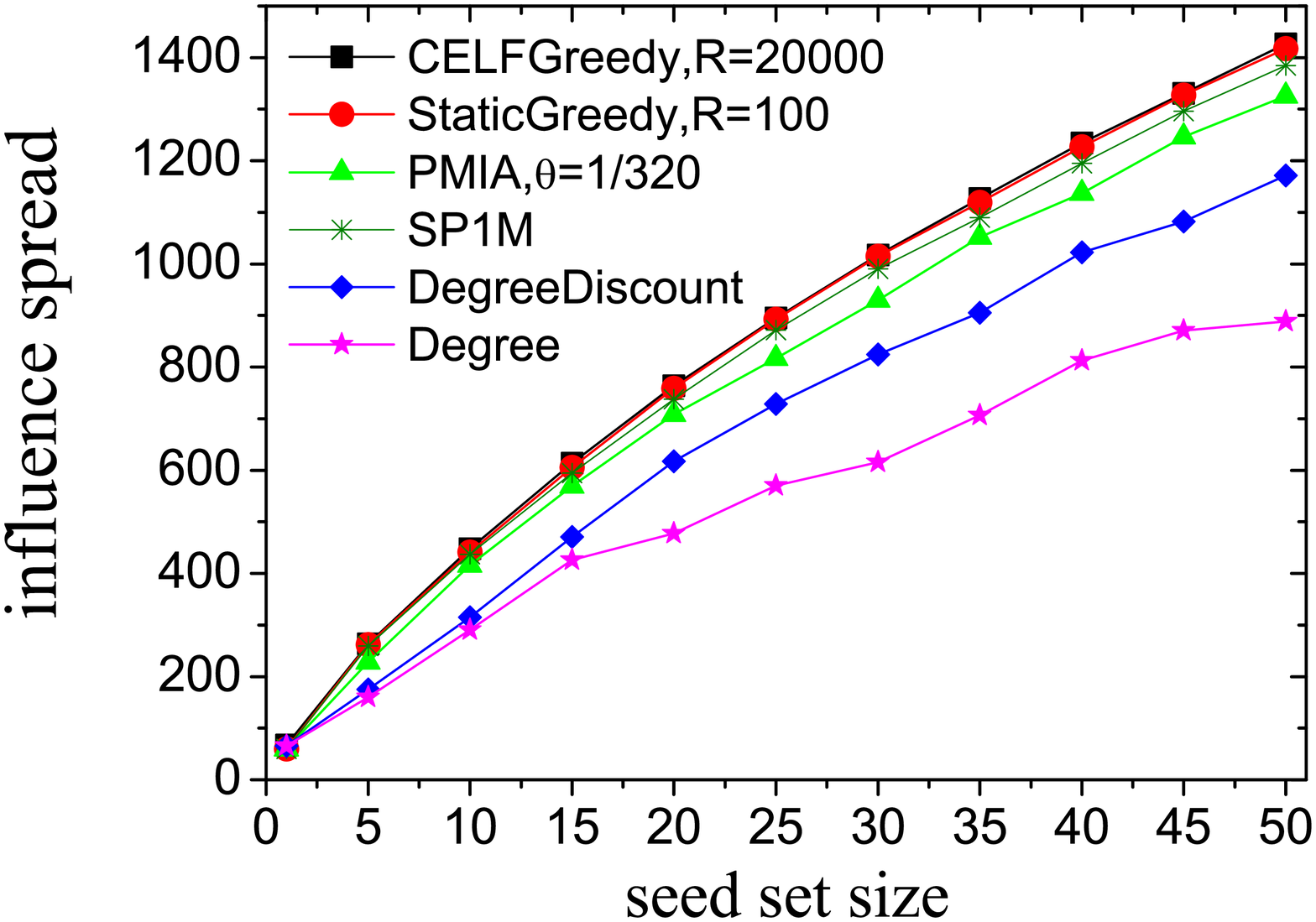}}
\subfigure[DBLP]
{\label{fig:subfig:DBLP} 
\includegraphics[width=0.31 \linewidth]{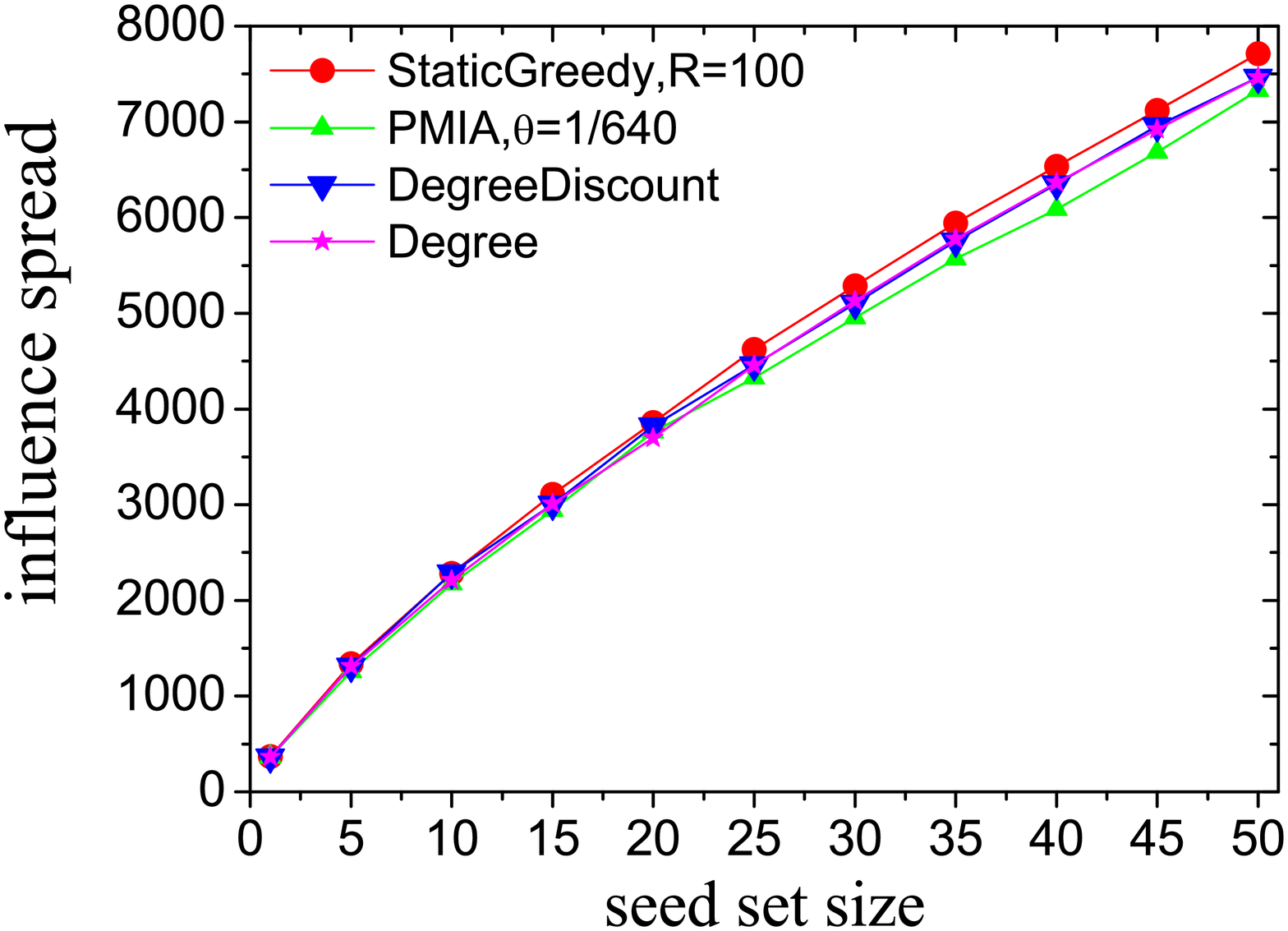}}
\subfigure[Epinions]
{\label{fig:subfig:NetHEPT} 
\includegraphics[width=0.31 \linewidth]{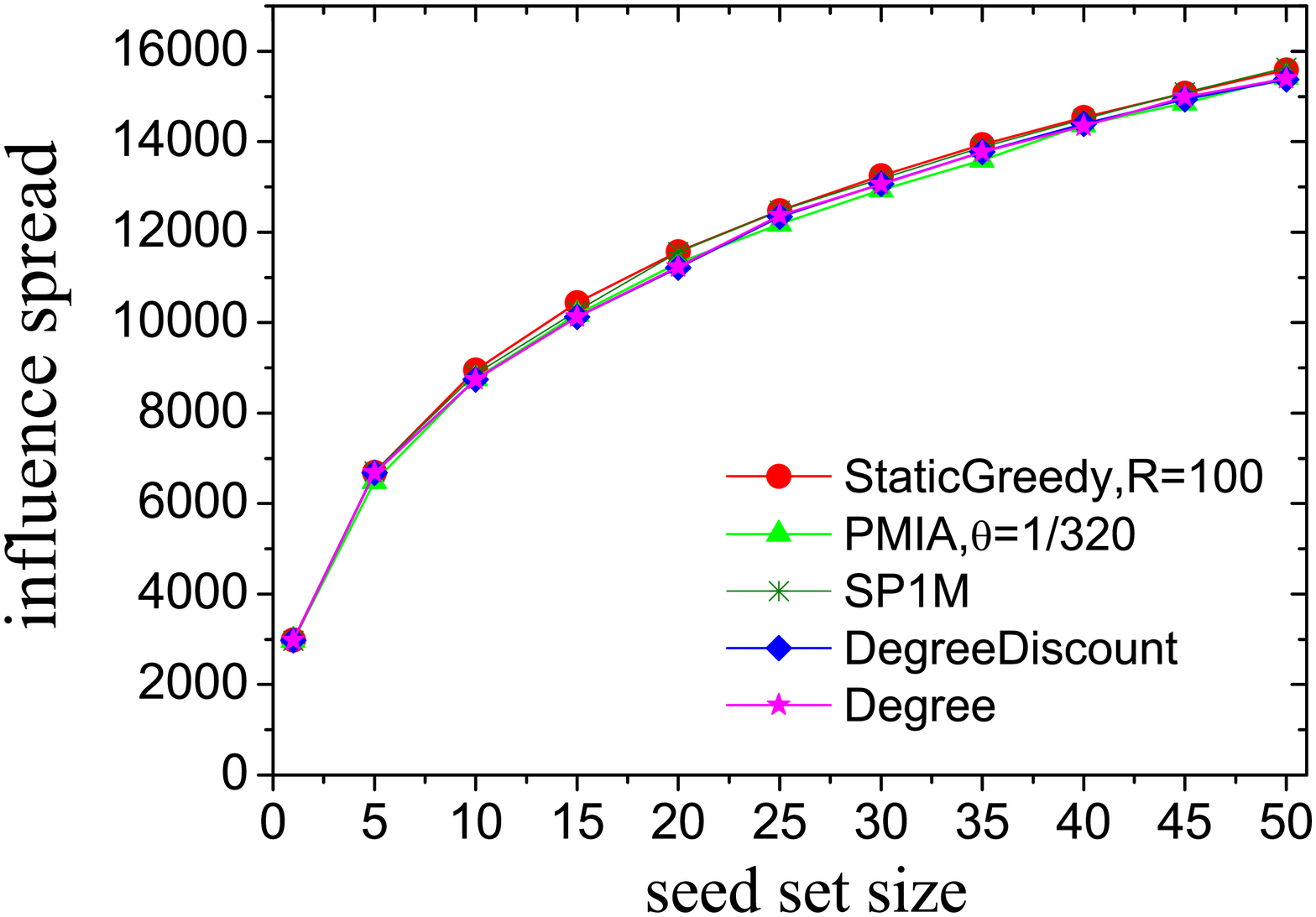}}
\label{fig:side:a}
\subfigure[Slashdot]
{\label{fig:subfig:NetPHY} 
\includegraphics[width=0.31 \linewidth]{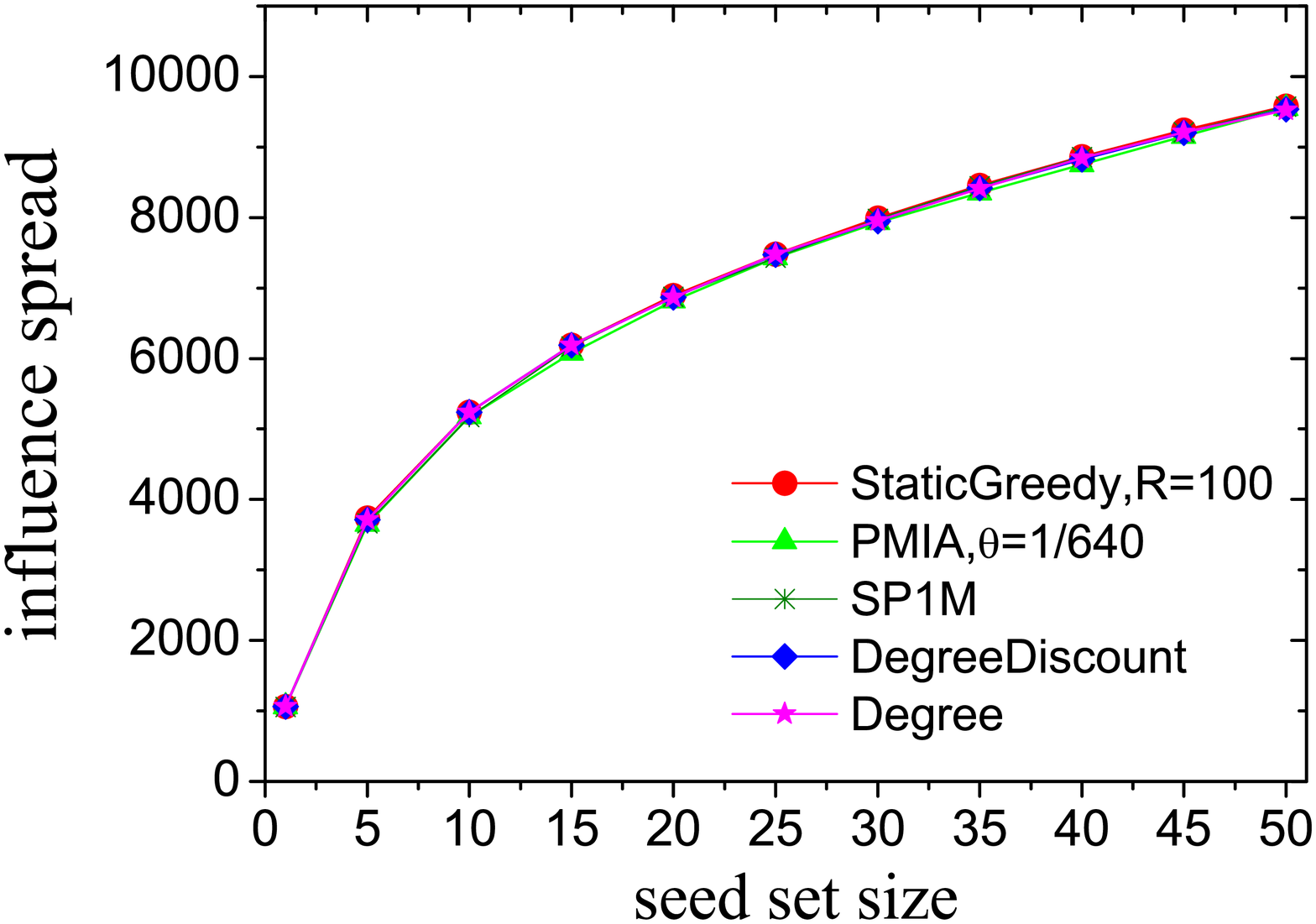}}
\subfigure[Douban]
{\label{fig:subfig:DBLP} 
\includegraphics[width=0.31 \linewidth]{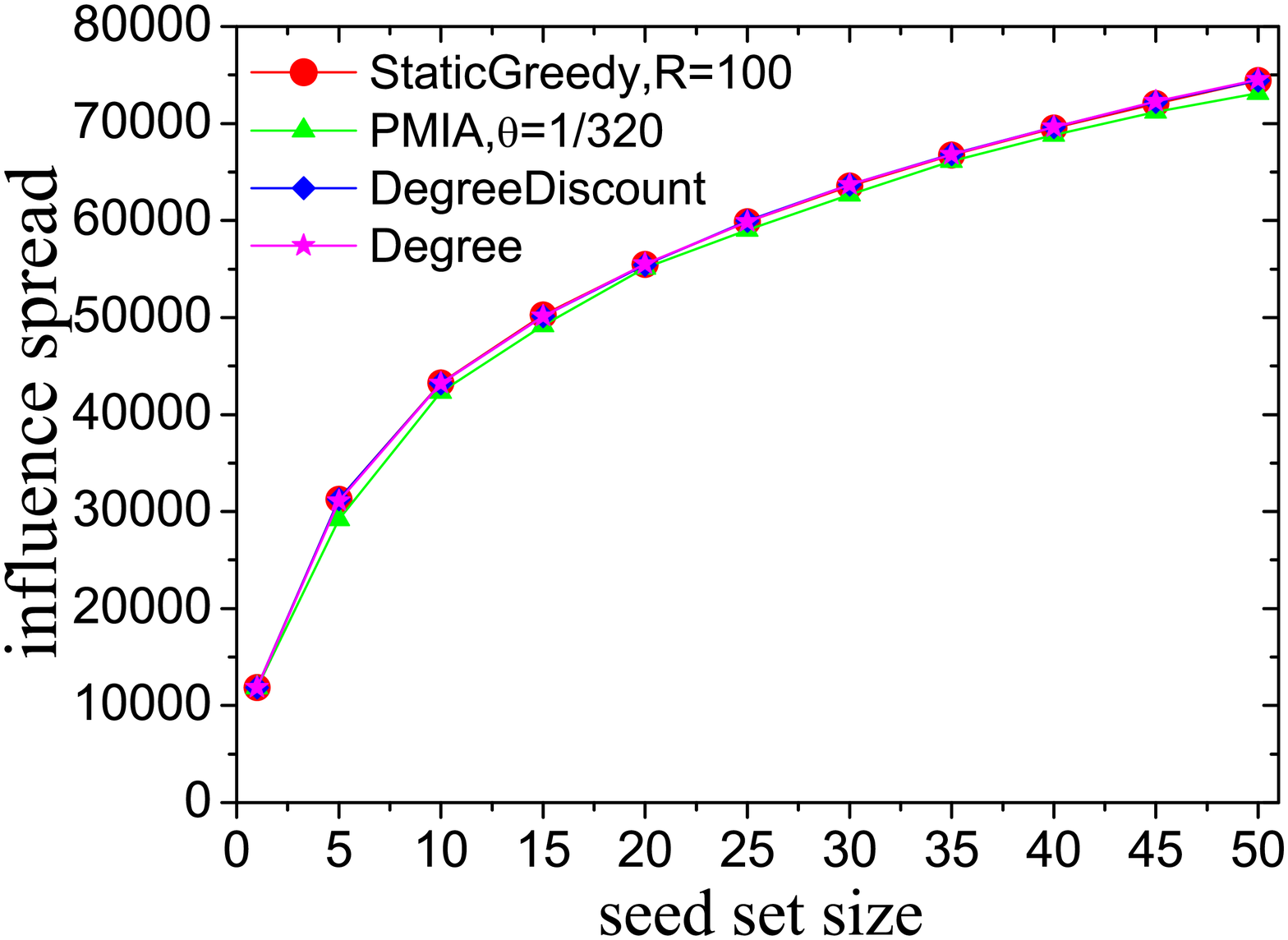}}
\caption{Influence spread under WIC model on six datasets.} \label{fig:performanceundirected.wic} 
\end{figure*}

We first compare the accuracy of the StaticGreedy algorithm with other algorithms by showing the influence spread of the obtained seed set. For every obtained seed set, $20,000$ Monte Carlo simulations are used to evaluate its influence spread. Figure~\ref{fig:performanceundirected.uic} shows the experimental results on influence spread for the six datasets under the UIC model. As shown in Figure~\ref{fig:performanceundirected:NetHEPT} and Figure~\ref{fig:performanceundirected:NetPHY}, the CELFGreedy algorithm provides the best influence spread on the moderate sized networks NetHEPT and NetPHY where the CELFGreedy algorithm is feasible to run. On the dataset NetHEPT, all the algorithms except the Degree heuristic algorithm have the influence spread similar to the CELFGreedy algorithm. However, on the dataset NetPHY, the differences among these algorithms become visible. StaticGreedy algorithm is still very close to the CELFGreedy algorithm and outperforms all the other algorithms. The difference between StaticGreedy algorithm and the CELFGreedy algorithm is less than $2\%$. Note that the accuracy of the StaticGreedy algorithm is obtained with a very small $R=100$ and can be further improved with larger $R$. For the rest networks with large scale where the CELFGreedy algorithm is infeasible, we compare the StaticGreedy algorithm with the other three baseline algorithms. We can see that the StaticGreedy algorithm always has the best accuracy. In particular, for the DBLP dataset, the StaticGreedy algorithm significantly outperforms the competing algorithms.
We further test StaticGreedy algorithm on the six test datasets with respect to the WIC model. For the moderate sized networks NetHEPT and NetPHY where CELFGreedy is feasible, as shown in Figure~\ref{fig:performanceundirected.wic:NetHEPT} and Figure~\ref{fig:performanceundirected.wic:NetPHY}, the StaticGreedy algorithm has almost the same influence spread to the CELFGreedy algorithm, which is the most accurate greedy algorithm. Moreover, StaticGreedy algorithm outperforms the other algorithms with a visible gap. For the DBLP, Epinions, Slashdot and Douban networks with large scale, the CELFGreedy algorithm is not scalable to these networks while StaticGreedy algorithm performs well. Moreover, for DBLP, Epinions networks, the StaticGreedy algorithm has slight higher accuracy than the other three baseline algorithms, and for Slashdot and Douban, it has consistent accuracy with the other algorithms. For these networks, due to their structural characteristics, a simple degree algorithm is good enough for influence maximization under the WIC model. However, for a given network, it is hard to determine a priori whether a simple heuristic is enough for influence maximization.

As demonstrated by the results on the test networks with both the UIC and WIC models, StaticGreedy algorithm has guaranteed accuracy as the original greedy algorithm and outperforms the state-of-the-art heuristics. Moreover, compared with the original greedy algorithm, the guaranteed accuracy of the StaticGreedy algorithm is obtained with the number of Monte Carlo simulations dramatically reduced by two orders of magnitude, i.e., from $20,000$ to $100$.

\subsubsection{Running time comparison}

We now test the running time of StaticGreedy algorithm and the competing algorithms. For StaticGreedy, we test the running time of both the StaticGreedyDU algorithm and the StaticGreedy algorithm with CELF optimization, denoted as StaticGreedyCELF. For heuristic algorithms, we test the running time of the PMIA and SP1M. We neglect the Degree and Degree discount algorithms since their accuracy are always lower than the PMIA and SP1M algorithms.

\begin{figure}[ht]
\centering
\subfigure[UIC model]{
    \label{fig:runningtime:UIC}
    \includegraphics[width = 0.42 \textwidth]{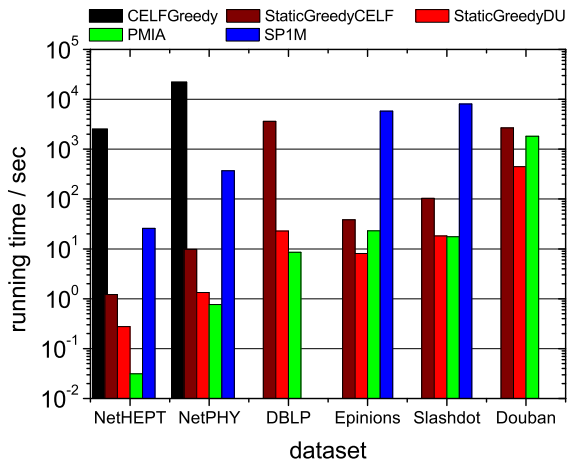}
}

\subfigure[WIC model]{
    \label{fig:runningtime:WIC}
    \includegraphics[width = 0.42 \textwidth]{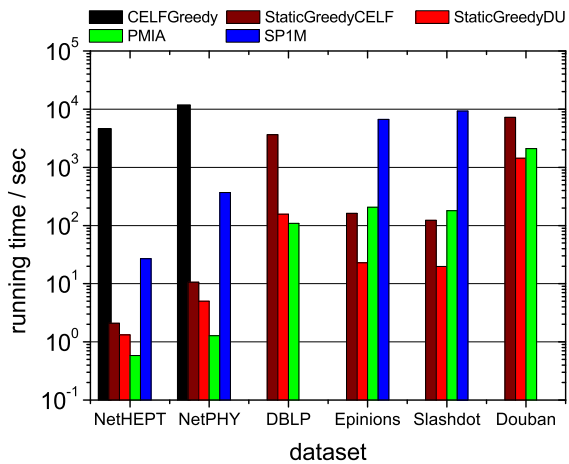}
}
\caption{\label{fig:runningtime}Running times of different algorithms on test
datasets under the UIC model and the WIC model.}
\end{figure}

Figure~\ref{fig:runningtime} shows the experimental results. For the six test networks, the StaticGreedyDU algorithm always runs 2-7 times faster than the StaticGreedyCELF algorithm, and the improvement is more significant for DBLP. The CELFGreedy algorithm is quite slow even for the moderate sized datasets, i.e., NetHEPT and NetPHY. The CELFGreedy algorithm requires several hours while our static greedy algorithms only take several seconds. The two static greedy algorithms reduce the running time by three orders of magnitude, compared with the CELFGreedy algorithm. More importantly, our static greedy algorithms obtain the reduction of running time without affecting the guaranteed accuracy. The time cost of our static greedy algorithms also significantly outperform the SP1M algorithm, which is not scalable and becomes infeasible to run for some large scale networks, such as DBLP and Douban networks. Furthermore, the running time of our static greedy algorithms is comparable to the PMIA algorithm, which is the most scalable heuristic algorithm. Note that the accuracy of the PMIA algorithm is unguaranteed. In addition, the StaticGreedyDU algorithm even outperforms the PMIA algorithm on three large scale networks, Epinions, Slashdot and Douban. It seems that our algorithm has the potential advantage on large scale networks compared with the PMIA algorithm.



\section{Conclusion and Future work}

In this paper, we have analyzed the scalability-accuracy dilemma of greedy algorithms for influence maximization, which has roots in the unguaranteed submodularity and monotonicity in existing implementations. We propose a static greedy algorithm to combat this problem by sharing Monte Carlo simulations in different iterations. Since both submodularity and monotonicity are strictly guaranteed, the static greedy algorithm always converges much more quickly than existing greedy algorithms. Hence, the proposed algorithm achieves the same accuracy with the state-of-the-art greedy algorithms while the number of Monte Carlo simulations needed is dramatically reduced by two orders of magnitude. We further give a dynamic update strategy taking advantage of the static snapshots to improve the static greedy algorithm, by applying which our algorithm becomes comparable to the most scalable heuristic algorithm. In addition, the idea behind the static greedy algorithm can be easily generalized to linear threshold model.

For the future work, we will study how to determine the minimum number $R$ of Monte Carlo simulations given network structure and diffusion model. Furthermore, we will implement the proposed static algorithm towards the frame of parallel computing to further improve the computational efficiency. We also look forward to seeing more applications of our algorithm on real world networks and practical scenarios. 

\section{Acknowledgments}
This work was funded by the National Natural Science Foundation of China with Nos 61202215, 61232010, 61202213, 61174152, and the National Basic Research Program of China (973 program) under grant number 2013CB329602. The authors thank Wei Chen for providing the codes of the PMIA algorithm and the datasets: NetHEPT, NetPHY and DBLP. The authors also thank to Jure Leskovec for providing the download for the social network datasets, Epinions and Slashdot. The authors thank to the member of the group NASC(www.groupnasc.org).


\bibliographystyle{abbrv}

\end{document}